# Reversible Graphene decoupling by NaCl photo-dissociation

I. Palacio[1,*], L. Aballe[2], M. Foerster[2], D. G. de Oteyza[3,4,5], M. García-Hernández[1], J.A. Martín-Gago[1*]

[1] *Materials Science Factory, Dept. Surfaces, Coatings and Molecular Astrophysics, Institute of Material Science of Madrid (ICMM-CSIC), C/Sor Juana Inés de la Cruz 3, 28049 Madrid, Spain*

[2] *ALBA Synchrotron, Carrer de la llum 2-26, Cerdanyola del Vallès, Barcelona 08290, Spain*

[3] *Donostia International Physics Center (DIPC), Paseo Manuel Lardizabal 4, 20018 San Sebastián, Spain*

[4] *Ikerbasque, Basque Foundation for Science, 48011 Bilbao, Spain*

[5] *Materials Physics Center, Centro de Física de Materiales (CSIC/UPV-EHU), Paseo Manuel Lardizabal 5, 20018 San Sebastián, Spain*

* Corresponding author e-mail address: i.palacio@csic.es, gago@icmm.csic.es

**ABSTRACT.**

We describe the reversible intercalation of Na under graphene on Ir(111) by photo-dissociation of a previously adsorbed NaCl overlayer. After room temperature evaporation, NaCl adsorbs on top of graphene forming a bilayer. With a combination of electron diffraction and photoemission techniques we demonstrate that the NaCl overlayer dissociates upon a short exposure to an X-ray beam. As a result, chlorine desorbs while sodium intercalates under the graphene, inducing an electronic decoupling from the underlying metal. Low energy electron diffraction shows the disappearance of the moiré pattern when Na intercalates between graphene and iridium. Analysis of the Na 2p core-level by X-ray photoelectron spectroscopy shows a chemical change from NaCl to metallic buried Na at the graphene/Ir interface. The intercalation-decoupling process leads to a n-doped graphene due to the charge transfer from the Na, as revealed by constant energy angle resolved X-ray photoemission maps. Moreover, the process is reversible by a mild annealing of the samples without damaging the graphene.

# 1. INTRODUCTION

Despite the exceptional potential of graphene (Gr) [1], the full control of its properties [2][3] in order to introduce improvements for future technological applications is still one of the main challenges in the field. In this respect, one of the most common approaches is the intercalation of atoms or molecules between graphene and the supporting substrate [4][5][6]. Actually, intercalation targets a two-fold objective. On the one hand, it may electronically decouple the graphene from the substrate [7][8][9][10][11] where it has been grown allowing for the recovery of its exceptional properties, usually degraded by the interaction with the substrate. On the other hand, the intercalation of molecules or atoms may lead to changes in the optical and electronic properties of graphene [12]. Regardless of the final goal, the different intercalation processes themselves merit comprehensive studies. Among the different works found in literature, the intercalation and adsorption of alkali metals such as potassium [13], lithium [14], cesium [15] or sodium [16][17][18] have been proven to be very appealing to engineer the band structure of graphene. In the case of sodium there is still controversy on whether it intercalates or adsorbs on top of graphene [16][17][18][19][20].

In this work we demonstrate a new and simple route for Na intercalation. It consists in the adsorption of a NaCl thin film on top of graphene and its subsequent photo-dissociation by irradiation with an X-Ray beam. We will show that chlorine desorbs immediately from the surface whereas sodium intercalates under graphene, decoupling it from the substrate. The process differs from other works where Na is directly evaporated onto the graphene surface. For instance, Watcharinyanon and coworkers [18][21] studied the adsorption and further intercalation of sodium by annealing on a Gr/SiC sample. This group also reported the intercalation of sodium by a soft x-ray exposure.

In our experiments we have used an epitaxially grown Gr on Ir(111) [22]. The intercalation can be followed by Low Energy Electron Diffraction (LEED), as the spots related to the moiré superstructure disappear. X-Ray spectroscopy (XPS) shows the vanishing of the Cl 2p core-level peak while Na 2p changes its chemical state from sodium in a salt crystal to metallic sodium. Moreover, Angle Resolved Photoemission (ARPES) maps reveals a shift of the Dirac point indicating a strongly n-doped graphene. The intercalation – decoupling process is reversible, and the Gr/Ir(111) sample can be easily recovered by annealing at 823 K.

## 2. RESULTS AND DISCUSSION

Our starting point (from now on, Stage 1 in the figures) is a NaCl film deposited on a Gr/Ir(111) sample. After a few seconds of an X-Ray beam exposure, the NaCl overlayer dissociates; chlorine desorbs and sodium intercalates, leading to a completely different system: Gr/Na/Ir(111) (in the following, Stage 2). At this stage the graphene is structurally decoupled from the surface, however upon annealing to 823 K, the Na is removed and a standard Gr/Ir(111) sample is recovered (in the following, Stage 3). X-ray measurements of the sample in Stage 1 were performed by illuminating and analyzing an area of about 400 μm$^2$ and moving to a fresh area once changes in the photoemission lineshape were detected. The photon flux is about $2.5 \times 10^9$ photons/s·μm$^2$.

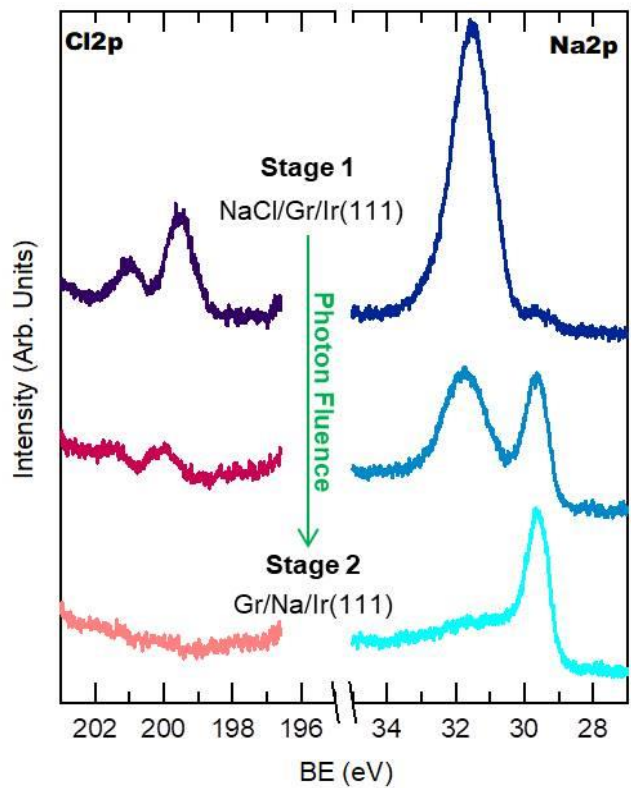

*Figure 1. Evolution of Cl 2p and Na 2p core levels of a NaCl/Gr/Ir(111) sample during XPS measurement as a consequence of the photon irradiation. The core level line-shape changes in a few seconds (from Stage 1 to Stage 2) when the substrate is irradiated with the photon beam. The photon*

*fluence increases from top to bottom. The photon energy is 260 eV and 136 eV for the Cl 2p and the Na 2p, respectively.*

Fig. 1 summarizes the changes of the Cl 2p and the Na 2p core levels in a NaCl/Gr/Ir(111) sample during XPS analysis induced by the irradiation with the photon beam. The sample was measured at a photon energy of 260 eV and 136 eV for the Cl 2p and the Na 2p, respectively. At the beginning (Stage 1), the sample is completely covered by a film of NaCl (Fig. 1, upper spectra). The Cl2p shows the typical spin-orbit splitting ($\Delta=1.6$ eV). With increasing photon fluence, and after a few seconds of irradiation, the Cl 2p peak strongly decreases in intensity. A similar intensity decrease is observed for the original Na 2p peak around 31.5 eV, concomitant with the appearance of a new feature at its low binding energy side (middle spectra). Finally (bottom spectra), the Cl 2p peak completely disappears and the Na 2p peak displays a single feature at ~29.6 eV (Stage 2). This behavior points to the complete photo-dissociation of the NaCl film as well as to the complete desorption of chlorine from the surface.

According to J.Y. Chung *et al.* [23] the Na 2p core level of NaCl grown by molecular beam epitaxy on Si (100) displays a single peak at ~32 eV. Therefore, the Na 2p peak found in Stage 1 at ~31.5 eV should be attributed to the binding energy of $Na^+$ bound to $Cl^-$ in the NaCl layer [23][24]. On the other hand, It is well known that the adsorption of alkali-metal atoms on transition metal surfaces modifies the alkali core-level photoemission spectra displaying three features which are related to surface, bulk and interface atoms [25]. In particular, the Na 2p spectrum measured during the deposition of Na on Pd (100) surfaces displays a single broad peak at a binding energy of around 30 eV for the first Na deposited layer. Further deposition of Na produces the appearance of a second feature shifted approximately 1 eV to the higher binding energies. These features are related to the interface and surface layers of atoms in a metallic environment, respectively. By increasing the Na deposition above the second layer a third feature appears at a binding energy in between both peaks which is attributed to a layer of bulk atoms between the surface and interface layers [25]. Therefore, the feature at ~29.6 eV observed in this work after photon irradiation (Stage 2) can be attributed to metallic Na atoms at the interface between Gr and Ir(111) [17][25][21]. These results clearly show that NaCl dissociates upon an incident photon beam, chlorine desorbs while sodium intercalates between graphene and the Ir (111) surface. As pointed out above, the Na 2p core level only shows one component related to interfacial sodium therefore indicating that unlike other alkali metals such as Cs [15] or K [26] on a Gr/Ir(111) sample, there is no coexistence of adsorbed and intercalated phases, but only intercalation once the sample has been irradiated. Moreover, the decrease in the Na 2p intensity when Na is intercalated relates to the signal attenuation by the graphene layer on top and is further proof for the Na intercalation. We have repeated the same measurements with a hemispherical analyzer and an Al-Kα monochromatic X-Ray source in the laboratory finding that NaCl does not dissociate and subsequently, Na does not intercalate through graphene. This could be due to the low flux (around $10^5$ photons/s·µm$^2$) in comparison with that of the Synchrotron ($10^9$ photons/s·µm$^2$).

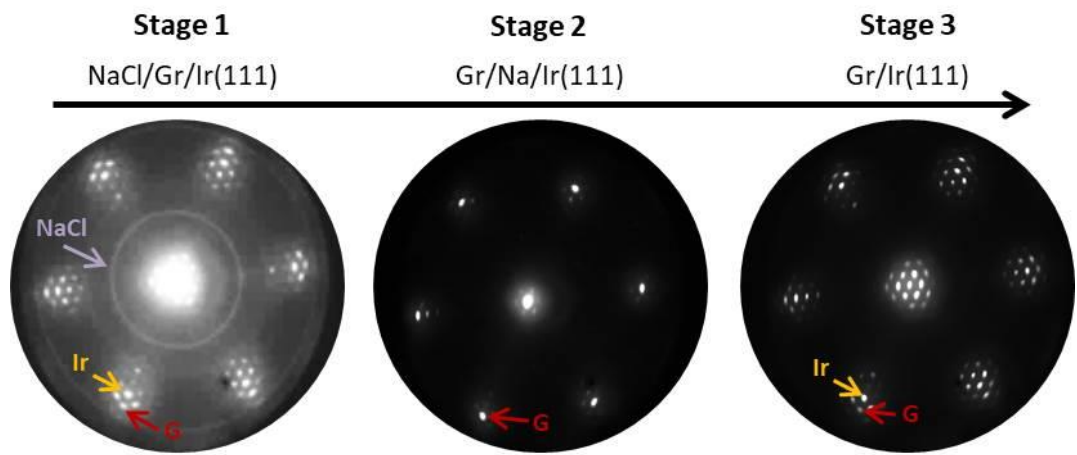

*Figure 2. Microspot-LEED patterns from stages 1 to 3: NaCl/Gr/Ir(111) → Gr/Na/Ir(111) → Gr/Ir(111) taken at 65 eV, 60 eV and 65 eV respectively.*

Additional structural information can be obtained using microspot-LEED. Fig. 2 shows the evolution of the LEED patterns of a NaCl/Gr/Ir(111) sample during the photo-dissociation of NaCl and the subsequent intercalation of Na. At Stage 1 (to be compared with Stage 1 of Fig. 1), the LEED pattern (taken at 65eV) corresponds to that of a NaCl/Gr/Ir(111) sample. The yellow arrow points to one of the Ir(111) spots and the red arrow points to one spot from the Gr lattice. The hexagons around the primary Ir spots are related to the growth of one monolayer of graphene on top of Ir (111) and correspond to the typical moiré pattern of Gr/Ir(111) [27]. In addition, the appearance of an inner ring is related to a multi-domain structure of the NaCl film [28]. When photo-dissociation of NaCl takes place, Stage 2, NaCl is completely dissociated: Cl$^-$ ions have desorbed and Na$^+$ ions have intercalated through graphene. At this stage, the ring structures related with NaCl have disappeared, as the salt has dissociated. On the other hand, the spots of the moiré structure, formed due to the mismatch of the Ir(111) and the Gr lattices, are almost inexistent, pointing out that sodium has decoupled the graphene from the substrate removing the superstructure. Finally, when the sample is annealed up to 823K (Fig. 2, Stage 3), the moiré related spots in the LEED pattern are recovered, indicating that all the intercalated sodium has been desorbed and the graphene is no longer decoupled from the metal substrate. The result is similar to that of a pristine Gr/Ir(111) sample.

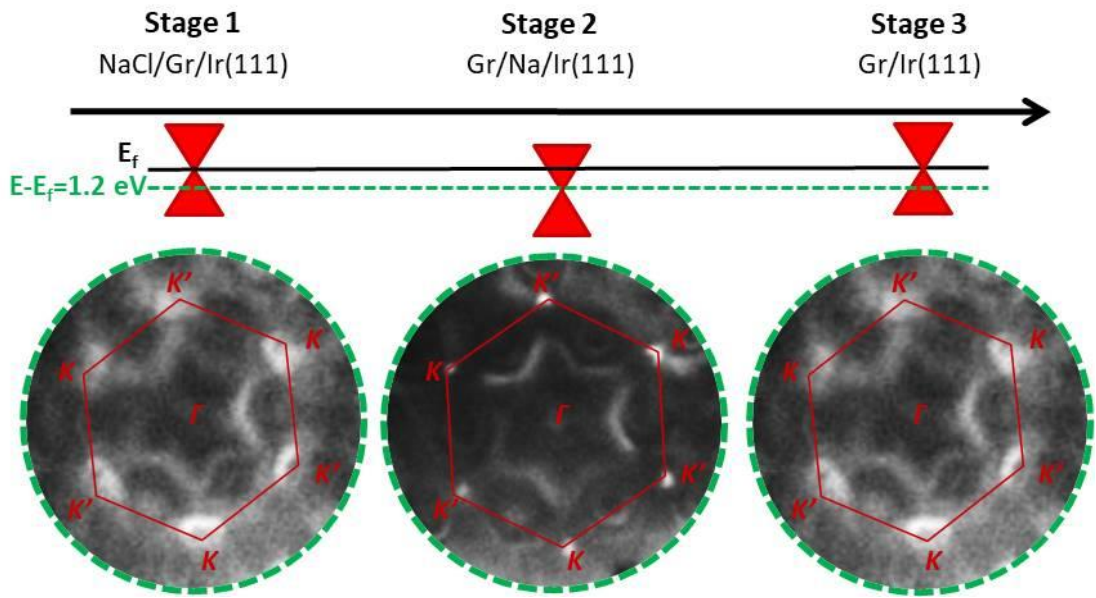

*Figure 3. Constant energy maps taken at $E-E_f=1.2$ eV. The red hexagon marks the first graphene Brillouin zone. From Stage 1 to Stage 2 the sample gets n-doped, whereas its initial properties are recovered after an annealing at 823 K (Stage 3).*

Finally, the evolution of the electronic structure of the NaCl/Gr/Ir(111) system when irradiated with photons was studied by means of microspot-ARPES. Due to the fast transformation of Stage 1 into Stage 2, full ARPES scans (k-maps) varying the energy cannot be obtained. Instead, constant energy

ARPES maps recorded at a binding energy of 1.2 eV are shown in Fig. 3, together with a schematic of the Dirac cone changes for an easy comprehension of the doping level at every stage. Several bands can be seen in every pattern, including the iridium bulk bands, but we exclusively focus on the dispersive π-bands (Dirac cones) of graphene. As a guide to the eye, the hexagon corresponding to the first graphene Brillouin zone, as well as points **K** and **K'** have been superimposed in red. Fig. 3 shows that a constant energy cut at $E-E_f$ = 1.2 eV of the NaCl/Gr/Ir(111) (Stage 1) system consists of hollow features centered around the K-K´ points, indicating the low doping of the system (Fig.3 left panel, Stage 1). Interestingly, the dispersive π-bands of the sample at Stage 1 are equal to those of a pristine Gr/Ir(111) (Stage 3), slightly p doped (0.1 eV) [29] [30], pointing out the weak interaction of the NaCl film with the graphene.

In contrast, in the equivalent constant energy map (taken at $E-E_f$=1.2 eV), taken after photo-dissociation of NaCl followed by sodium intercalation, the dispersive π-bands appear as spots, implying a substantial downward shift of the Dirac cone corresponding to strongly n-doped graphene. This is shown in Fig. 3 middle panel (Stage 2). Pervan *et al*. [16] have already reported a shift of the Dirac Point to around 1.2-1.3 eV below the Fermi level when Na is intercalated on a Gr/Ir(111) sample, indicating an efficient mechanism of charge transfer from the alkali metal to the graphene. The controversy about whether sodium intercalates or not underneath graphene at RT [16], [20] [31] does not apply in our scenario, since we evaporate NaCl, and dissociate it with photons. A n-doping have already been reported by Papagno *et al*.[20] and Jeon *et al*. [19] for a slightly different system in which Na is adsorbed on top of graphene. Nonetheless, a similar behavior can be expected when Na is intercalated. Lastly, and after annealing the sample at 823 K (Stage 3), sodium desorbs and the energy map of the system recovers its initial state, as it is shown on the right panel of Fig. 3, where at 1.2 eV binding energy, the cut of the π-bands is again hollow, demonstrating that the sample is coupled again with the substrate and is not n-doped anymore. These results also suggest that the graphene quality is not altered after the process of intercalation/desorption of the Na.

Although it is clear from the above presented experiments that Na intercalates through graphene, the mechanisms leading to this process are not completely understood. Density functional theory calculations [31], as well as other works [32] [15] [17], indicate that intercalation is favourable over adsorption. Although the precise mechanism of intercalation is still under debate, it is nowadays accepted that surface defects (ie: grain boundaries, edges, wrinkles crossing or nanobbubles) are efficient intercalation channels through the graphene layer [9] [15][17][32].

## 3. CONCLUSIONS

In this work we have shown that it is possible to electronically decouple graphene from Ir(111) by intercalation of Na. The intercalation process takes places after photo-dissociation of an overlayer of NaCl that has been previously deposited on the graphene surface. NaCl dissociates upon a short photon exposure and chlorine desorbs while sodium intercalates through graphene, decoupling it from the metal substrate. This process can be easily followed by microspot x-ray spectroscopy and low energy electron diffraction. Moreover, the electronic structure of the system studied by means of ARPES shows a strongly n-doped character of graphene when the decoupling is complete. The whole process is reversible through a soft annealing without damaging the graphene.

## 4. METHODS

The experiments have been carried out at the PEEM experimental station of the CIRCE beamline at the ALBA Synchrotron [33]. All measurements were done in a low energy and photoemission electron microscope and thus all data are obtained from micrometer-sized regions. Since the sample surface is homogeneous, we mostly used the largest available aperture size of 10 μm diameter, although sometimes we used 5 μm diameter. Samples were prepared in an ultra-high vacuum (UHV) chamber with a base pressure of $1 \times 10^{-10}$ mbar. NaCl (99.9% Sigma Aldrich) was sublimated from a homemade Ta crucible annealed at 803 K controlled by a type-K thermocouple spot-welded to it with the sample at room temperature. Ir(111) surfaces were cleaned by repeated cycles of argon ion sputtering and annealing in an oxygen atmosphere (T = 1373 K and $P_{oxygen}$= 2×10$^{-8}$ mbar). In order to avoid any residual oxygen on the surface, several extra cleaning cycles were carried out without oxygen. Graphene was grown on Ir(111) in a decomposition process of ethylene. In a first step, ethylene (P = 1×10$^{-8}$ mbar during 30 seconds) is adsorbed on the sample at room temperature (RT), then the sample is flashed up to 1373 K for 30 s. In a second step the sample is exposed to higher ethylene pressure (P = 1×10$^{-7}$ mbar), again followed by a thermal decomposition at 1373 K for 7 min [22].

**ACKNOWLEDGEMENTS**

This work has been supported by the EU Graphene Flagship funding (Grant Graphene Core2 785219), the Spanish MINECO (MAT2017-85089-C2-1R) and the EU via the ERC-Synergy Program (Grant ERC-2013-SYG-610256 NANOCOSMOS). We are grateful to Prof. C. Palacio for their fruitful discussions about the presented work.